\documentclass[pra,twocolumn,showpacs]{revtex4}
\usepackage[dvips]{graphicx}
\usepackage{graphicx,color}

\begin{document}

\title{Entanglement from density measurements: analytical density-functional for the entanglement of strongly correlated fermions}

\author{Vivian V. Fran\c{c}a}\email{vivian.franca@physik.uni-freiburg.de}
\affiliation{Capes Foundation, Ministry of Education of Brazil, CP 250, 70040-020, Brasilia, Brazil}
\affiliation{Physikalisches Institut, Albert-Ludwigs-Universit\"at, Hermann-Herder-Str. 3, Freiburg, Germany}

\author{Irene D'Amico}\email{irene.damico@york.ac.uk}
\affiliation{Department of Physics, University of York, York YO10 5DD, United Kingdom}

\begin{abstract}

We derive an analytical density functional for the single-site entanglement of the one-dimensional homogeneous Hubbard model, by means of an approximation to the linear entropy. We show that this very simple density functional reproduces quantitatively the exact results. We then use this functional as input for a local density approximation to the single-site entanglement of inhomogeneous systems. We illustrate the power of this approach in a harmonically confined system, which could simulate recent experiments with ultracold atoms in optical lattices as well as in a superlattice and in an impurity system. The impressive quantitative agreement with numerical calculations -- which includes reproducing subtle signatures of the particle density stages -- shows that our density-functional can provide entanglement calculations for actual experiments via density measurements. Next we use our functional to calculate the entanglement in disordered systems. We find that, at contrast with the expectation that disorder destroys the entanglement, there exist regimes for which the entanglement remains almost unaffected by the presence of disordered impurities.

\end{abstract}


\maketitle

\newcommand{\be}{\begin{equation}}
\newcommand{\ee}{\end{equation}}
\newcommand{\bea}{\begin{eqnarray}}
\newcommand{\eea}{\end{eqnarray}}
\newcommand{\bi}{\bibitem}
\newcommand{\la}{\langle}
\newcommand{\ra}{\rangle}
\newcommand{\ua}{\uparrow}
\newcommand{\da}{\downarrow}
\renewcommand{\r}{({\bf r})}
\newcommand{\rp}{({\bf r'})}
\newcommand{\rpp}{({\bf r''})}

\section{Introduction} 

Entanglement, one of the most intriguing features of quantum mechanics, is an essential ingredient in quantum information theory \cite{horodecki}, and it is also receiving increasing attention in the study of fundamental phenomena such as quantum phase transitions (QPT) \cite{rev.solids}. Entanglement has been investigated in many different physical systems, such as photons in optical cavities \cite{davi}, ultracold atoms interacting with light \cite{vvf2007, porras}, and in solid state systems \cite{rev.solids}, where it has been associated to particle-particle correlations \cite{plenio}.

Recently, special attention has been devoted to solids, since they are considered one of the most promising systems for the development of quantum information devices \cite{stoneham}. In particular, the Hubbard model \cite{hubb}, which captures most important physics of interacting many-body systems \cite{tasaki, paiva, machida, derz}, has been intensively used to understand entanglement in solids \cite{gu1, larsson1, vvf2006, vvf2008, carr, coe2010,epl2010}. In this context, creating, distributing, quantifying, and controlling entanglement represent great challenges. Much progress has been achieved and some well defined entanglement measures, such as entropies and concurrence, have been proposed \cite{rev.solids}.
The von Neumann entropy has been used for measuring the bipartite entanglement of pure states \cite{rev.solids}. For systems with very many degrees of freedom, the linear entropy -- which gives an indication of the number and spread of terms in the Schmidt decomposition of the state -- has been proposed as a viable alternative~\cite{buscemi2007,irene2008,shirwan2009}.

For the homogeneous Hubbard model, the single-site entanglement \cite{zanardi} and its properties have been investigated \cite{vvf2006} and associated to QPT's \cite{oster, gu1, larsson1}. A local-density approximation (LDA) for the entanglement entropy of any generic inhomogeneous system has been proposed \cite{vvf2008} and applied to the Hubbard model for different types of inhomogeneities. Also block-block entanglement has been investigated in connection to QPT \cite{gu2} and from the view point of its universal and non-universal contributions \cite{vvf2008pra}. 

Another extremely important -- but less addressed -- question concerning entanglement quantifiers, is whether and how theoretical predictions can be connected to quantities that are measured experimentally. It has been shown that magnetic susceptibility can be an entanglement witness \cite{gosh, vedral, almeida}, which is used to point out the presence of entanglement. Also momentum-momentum correlation \cite{lukin, flo}, energies \cite{irene2008,shirwan2009}, and probability density at highly symmetric points~\cite{shirwan2009}
 have been proposed as entanglement indicators, which reproduce the general trend of entanglement. However, a quantity able to predict entanglement {\it quantitatively} and, in addition, with the potential of being {\it experimentally measurable}, is still missing. Here we show that a very promising candidate for this is the particle density. 
 
As a direct consequence of the Hohenberg-Kohn theorem \cite{hk}, the pillar of Density Functional Theory (DFT),
 ground-state entanglement must be a functional of the density. For the one-dimensional homogeneous Hubbard model this functional may be numerically accessed \cite{ba}, but an explicit expression cannot be obtained due to the implicit dependence of the ground-state energy on the density. Different analytical and numerical techniques have been proposed for this calculation \cite{gu1, larsson1, vvf2006}. Nevertheless, obtaining numerically the ground-state energy has been so far the main limiting factor for a complete analytical understanding of the entanglement in the homogeneous Hubbard model. For inhomogeneous systems the scenario is even more complicated, despite the application of powerful DFT methods \cite{vvf2008}. 

From the experimental point of view, it has been recently shown that the {\it measured} density in systems of ultracold atoms in optical lattices with single-site resolution \cite{greiner3, bloch2} can be very accurately reproduced by the single-site density of the Hubbard model. Several theoretically predicted phenomena have already been verified in such experiments, e.g., the superfluid to Mott-insulator transition \cite{greiner2} and BEC-BCS crossovers \cite{bart}; via DFT, the density may now become a powerful tool for measuring entanglement in experiments.

Motivated by this, in the present work we provide an \textit{analytical} expression for the density-functional for the entanglement of homogeneous strongly correlated systems. Our approach is based on an approximation of the linear entropy for the one-dimensional Hubbard model valid, in principle, for a specific range of interactions and densities. The functional is then extended to a vast range of parameters by means of well established transformations. We test the performance of our density functional in homogeneous and, via an LDA, in three different inhomogeneous systems. We find that the entanglement in many-body systems can be \textit{quantitatively} obtained directly from our very simple density functional. Finally we use our method for investigating the entanglement in disordered systems. To achieve a good statistics, this analysis requires many realizations and, therefore, could not be explored so far, while our functional greatly reduces its computational cost. We find that in general disorder destroys entanglement. Surprisingly though, there exist regimes for which the degree of entanglement remains almost unaffected. 

\section{Analytic density functional for the homogeneous system}

We start from the homogeneous Hubbard model, 
\be
\hat{H}=-\sum_{i\sigma} \left(\hat{c}^\dagger_{i\sigma}\hat{c}_{i+1,\sigma}+H.c.\right)+u\sum_i \hat{n}_{i\uparrow}\hat{n}_{i\downarrow}+V\sum_{i\sigma} \hat{n}_{i\sigma},\label{hubb_h}
\ee
in which all sites are equivalent. Here $u$, the on-site interaction, and $V$, the external potential, are rescaled by the hopping parameter between neighboring sites; $\hat{c}^\dagger_{i\sigma}$ and $\hat{c}_{i\sigma}$ are fermionic creation and annihilation operators, and $\hat{n}_{i\sigma}=\hat{c}^\dagger_{i\sigma}\hat{c}_{i\sigma}$ is the number operator at site $i$ with spin $\sigma=\uparrow,\downarrow$.

\begin{figure}[!t]
\includegraphics[width=9cm]{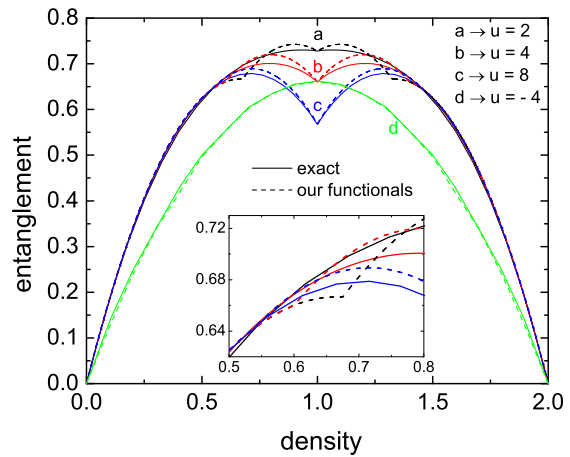}
\caption{(Color online) Single-site entanglement as a function of density for the homogeneous Hubbard model in the thermodynamic limit. For repulsive interactions ($u=2,4,8$) the exact results (solid lines a, b, and c) are compared to Eq.(\ref{lin2}) (dashed lines a, b, and c). For attractive interactions ($u=-4$) the exact results (solid line d) are compared to Eq.(\ref{uneg}) (dashed line d). {\it Inset}: zoom of main panel for $n\approx \alpha(u)+1/2$. The exact results were obtained from Eq.(\ref{linear}).} \label{hom}
\end{figure} 

The single-site entanglement is quantified using the linear entropy
\be
L_i=1-Tr(\rho_i^2),
\ee
where the reduced density matrix of site $i$, $\rho_i=Tr_B \rho$, is the trace of the total density matrix $\rho$ with respect to the remaining sites. 
For the homogeneous Hubbard model $\rho_i=diag\{w_\uparrow,w_\downarrow,w_2,w_0\}$, with $w_\sigma=\left\langle \hat{n}_\sigma\right\rangle-w_2$ the single spin occupation probability, $w_2=\left\langle \hat{n}_\uparrow \hat{n}_\downarrow \right\rangle$ the double occupation, and $w_0=1-w_\uparrow -w_\downarrow -w_2$ the zero occupation probabilities \cite{zanardi}.

This allows us to express the single-site entanglement of the homogeneous Hubbard model as an explicit function of the occupation probabilities
\be
L^{hom}=1-w_\uparrow^2-w_\downarrow^2-w_2^2-w_0^2.\label{linear}
\ee 
For non-magnetic systems the probabilities are given by
\bea
w_2&=&\frac{\partial e_0(n,u)}{\partial u}\label{w2}\\
w_\uparrow &=&w_\downarrow=\frac{n}{2}-w_2\label{wup}\\
w_0&=&1-n+w_2\label{w0},
\eea
where $e_0(n,u)$ is the per-site ground-state energy. 

In order to express the entanglement as an {\it explicit} function of the density, we will approximate Eq.(\ref{w2}). Initially, we focus on repulsive interaction and
$n<1$, but as shown later, the results will be extended to $n>1$ and to $u<0$ by means of particle-hole transformations. An exact analytical expression for the energy, and consequently for $w_2$, is unknown in this regime, except at $n=1$, for which the exact Lieb-Wu solution \cite{ba} gives 
\be
w_2(n=1,u>0)\equiv\alpha(u)=2\int_0^\infty{\frac{J_0(x)J_1(x)e^{ux/2}}{(1+e^{ux/2})^2}dx},\label{w2n1}
\ee
where $J_k(x)$ are Bessel functions of order $k$. 

For low-density systems the double occupancy is extremely reduced due to the repulsive interactions \cite{w2}; for higher densities, $w_2$ increases and can not be discarded. We then consider $w_2\approx 0$ when $L^{hom}(n,u>0)|_{w_2=0}\ge L^{hom}(n,u>0)|_{w_2=\alpha(u)}$, and assume Eq.(\ref{w2n1}) as an approximation for $w_2$ otherwise \cite{foot1}.

Then the linear entropy can be written as the {\it explicit} density functional  
\bea
L^{hom}(n,u>0)&\approx& 2n-\frac{3n^2}{2}+\label{lin2}\\
&&\hspace{-1.3cm}\left[(4n-2)\alpha(u)-4\alpha(u)^2\right]\Theta(n-\alpha(u)-1/2),\nonumber
\eea%
where $\Theta(x)$ is a step function with $\Theta(x)=0$ for $x<0$ and $\Theta(x)=1$ for $x\geq0$. We extend our functional to the regime of $n>1$ using the particle-hole symmetry, by replacing $n$ with $2-n$ in Eq.(\ref{lin2}). Such extension is necessary to describe a larger variety of systems, and also in view of extending the results via DFT methods to more realistic inhomogeneous systems.

\begin{figure}[t!]
\centering
\includegraphics[width=8.2cm]{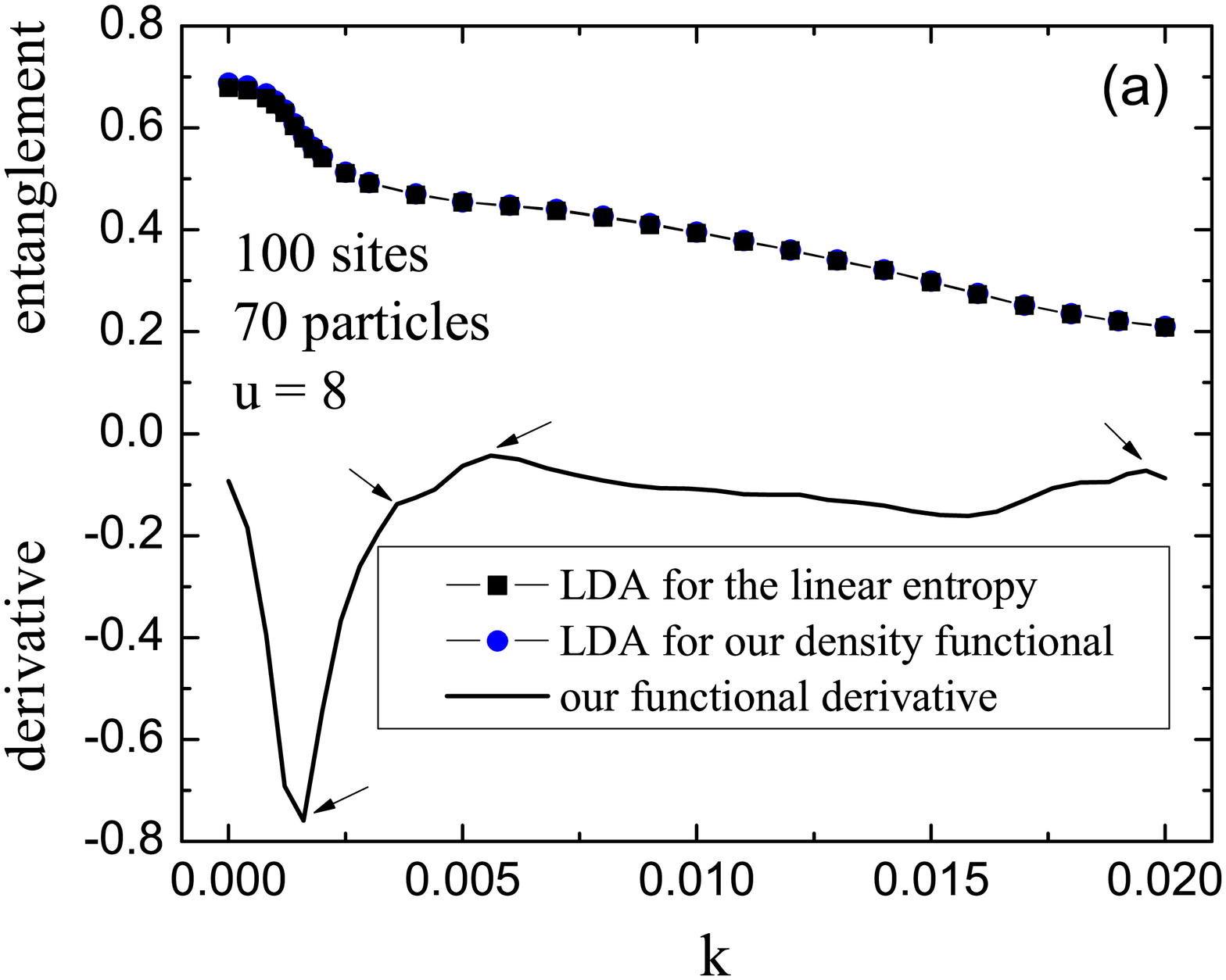}
\includegraphics[width=8.2cm]{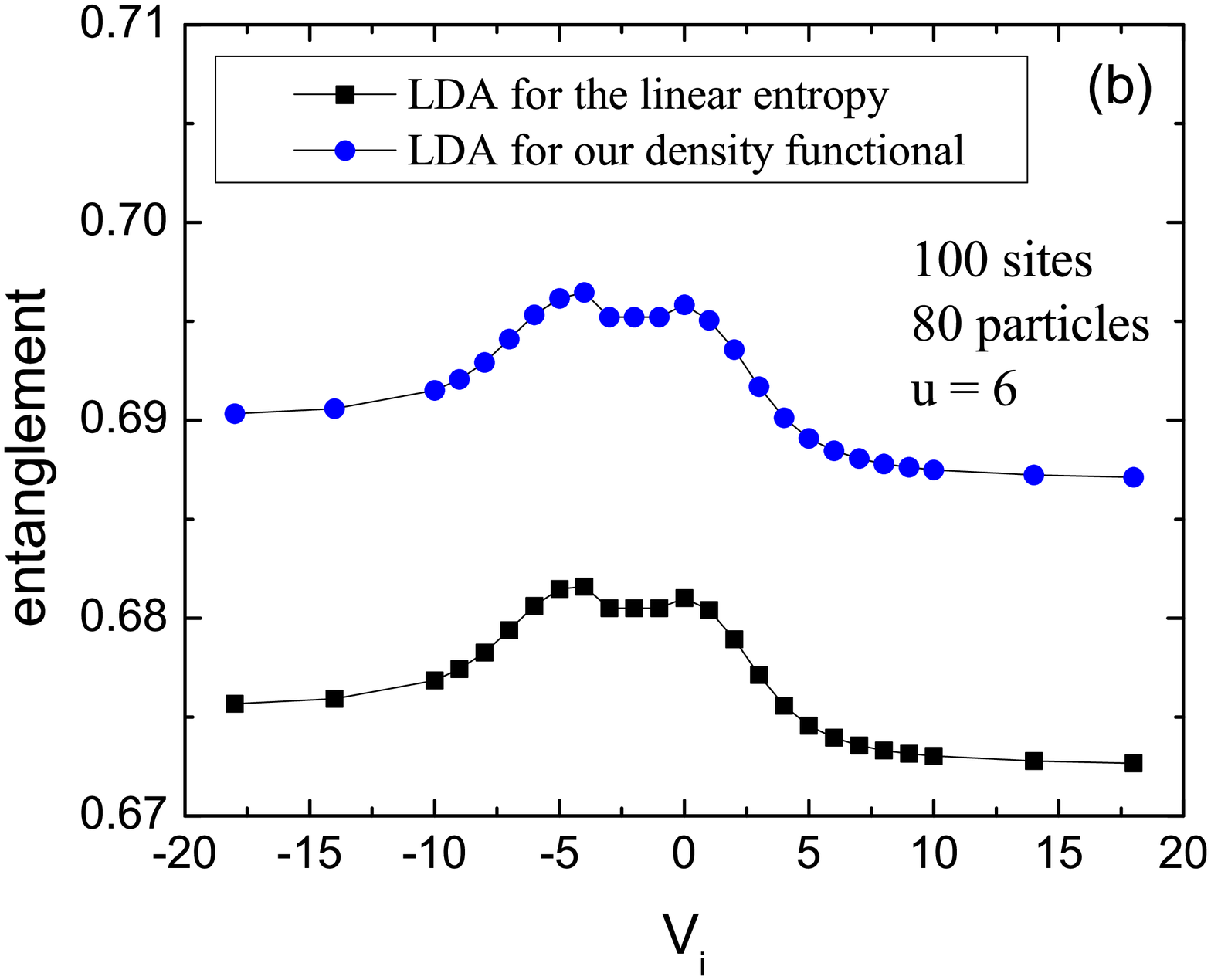}
\includegraphics[width=8.2cm]{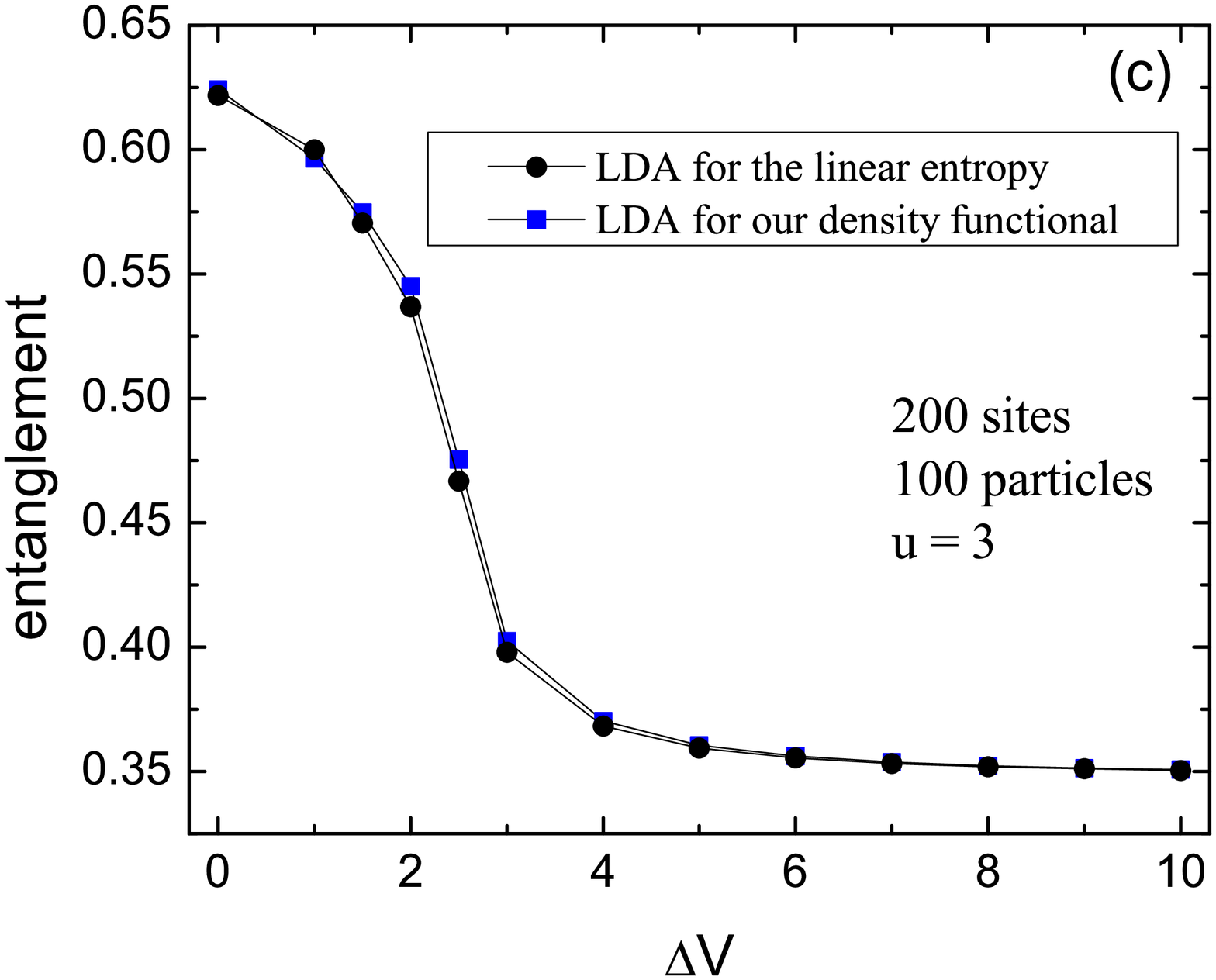}
\caption{Single-site entanglement as a function of the inhomogeneity: a) $k$, the curvature of the harmonic confinement, $V_i=k(i-i_0)^2$, with $i_0=N/2+1$, $N=100$, $u=8$, and $n=0.7$; the solid line is the derivative $\partial L^{LDA}/ \partial k$ from Eq.(\ref{linlda}) with Eq.(\ref{lin2}) as input and the arrows indicate stages of the density profile; b) $\Delta V$, the modulation amplitude of a superlattice with periodicity of $20$ lattice sites, for $N=200$, $100$ particles and fixed interaction $u=3$; c) $V_i$, the intensity of a single-impurity localized at the site $i=N/2+1$, for $N=100$, $80$ particles and $u=6$.} \label{harm}
\end{figure}

For attractive interactions the double occupancy can neither be discarded nor approximated by the Lieb-Wu result. The extension of our analysis to $u<0$ can be though obtained by means of the mapping between the energies of repulsive and attractive systems \cite{campo}, 
\be
e_0(n,u<0)\approx\frac{u n}{2}+e_0(n=1,u>0)\sin\left(\frac{\pi n}{2}\right),\label{map}
\ee
which is valid for any density. From (\ref{w2}) and (\ref{map}) we obtain 
\be
w_2(n,u<0)\approx\frac{n}{2}-\alpha(\left|u\right|)\sin\left(\frac{\pi n}{2}\right)\label{w2upos},
\ee
and the density-functional for the entanglement becomes
\bea
L^{hom}(n,u<0)&\approx& n-\frac{n^2}{2}+2\alpha(\left|u\right|)\sin\left(\frac{\pi n}{2}\right)\nonumber\\
&&-4\alpha^2(\left|u\right|)\sin^2\left(\frac{\pi n}{2}\right).\label{uneg}
\eea

In Figure \ref{hom} we compare our density-functionals for the entanglement, Eq.(\ref{lin2}) and Eq.(\ref{uneg}), to the exact entanglement Eq.(\ref{linear}). The agreement is very good, considering the simplicity of our density-functional expressions. For $u<0$ the exact results are {\it very accurately} reproduced because the double occupation is fully taken into account in Eq.(\ref{uneg}). For $u>0$ the results reveal that the relevant dependency on $n$ for the entanglement is taken into account by the other occupation probabilities, i.e., the main contribution of the double occupancy to the entanglement is almost independent on the density, therefore the simplified form of $w_2$ adopted in Eq.(\ref{lin2}) is enough for predicting the entanglement with good precision.
In particular Figure \ref{hom} shows that there are ranges of densities, $0\leq n\lesssim 0.5$ and $1.5\lesssim n\leq 2$, for which many-body interactions do not play a relevant role: in a homogeneous system, up to $n\approx0.5$ (and for $n\gtrsim1.5$ when considering particle-hole transformation), the double occupation is in fact strongly reduced as here $w_2\lesssim n^2/4$ ($w_2\lesssim (2-n)^2/4$). In turn in the Hubbard model, where only on-site interactions are considered, this implies a lesser role of the particle-particle interactions \cite{note_many_body}.

However there is a small range of combined parameters -- namely $n\approx \alpha(u)+1/2$ for $0<u\leq2$ (see inset of Fig. \ref{hom}) -- for which our approximation to $w_2$ is not as good: here also because on-site repulsion is small, $w_2$ may become significant. The {\it maximum} error in this specific region decays exponentially from $\sim 12\%$ at $u=0.5$ to $4.6\%$ at $u=2$. Considering that typical interactions in solids correspond to $u\sim6$ \cite{u.solids} and in experiments with optical lattices the interaction can be easily tuned in a wide range (up to $u\sim 200$) \cite{u.opt}, for most systems of practical interest our approach can provide the entanglement with a precision better than $\sim 3\%$ for any density. 

The very good agreement of our density functionals with exact calculations shows that the entanglement of actual experiments described by the Hubbard model~\cite{greiner3, bloch2} could be obtained accurately and {\it directly} from density measurements. Also our results imply much simpler calculations which opens possibilities for the entanglement investigation in more complex systems, such as the disordered media we will explore in the last section.

\section{Density Functional for Inhomogeneous Systems}

In order to investigate the performance of our density functional in inhomogeneous systems, we present in what follows numerical calculations for the Hubbard model with an inhomogeneous external potential $\sum_{i,\sigma}V_i\hat{n}_{i,\sigma}$. We adapt to this purpose the LDA formalism developed in Ref.\cite{vvf2008}. The average single-site linear entropy of inhomogeneous systems is then given by
\be
L^{inh}(n,u)\approx L^{LDA}(n,u)=\frac{1}{N} \sum_i^N L^{hom}(n,u)|_{n=n_i}.\label{linlda}
\ee
In Eq.(\ref{linlda}) the inputs are our density-functionals, Eqs.(\ref{lin2}) or (\ref{uneg}), calculated at the local $i$-site density $n_i$, and averaged over the $N$ sites \cite{foot2}.

We start with the inhomogeneous harmonic potential, which may model, for example, the trap in systems of ultracold atoms in optical lattices \cite{greiner3} or quantum dots \cite{quantumdot, coe2010, epl2010}. This inhomogeneous system contains very interesting physics: it undergoes subtle transitions in the density profile as the potential increases \cite{opticaltrap} and the single-site entanglement is able to point out each and every one of these phases \cite{vvf2008}. We have chosen $n=0.7$, which represents a typical density reached in the experiments \cite{morsch}, and $u=8$, as experimentally the interaction can be tuned up to $u\sim 200$ \cite{u.opt}. In Figure \ref{harm}--a we compare our density-functional for the entanglement to accurate numerical results. The agreement is indeed very good and all the stages of the particle density \cite{vvf2008} are correctly pointed out by the derivative of our functional (see the entanglement derivative in Fig.\ref{harm}--a). 

\begin{figure}[!t]
\includegraphics[width=9cm]{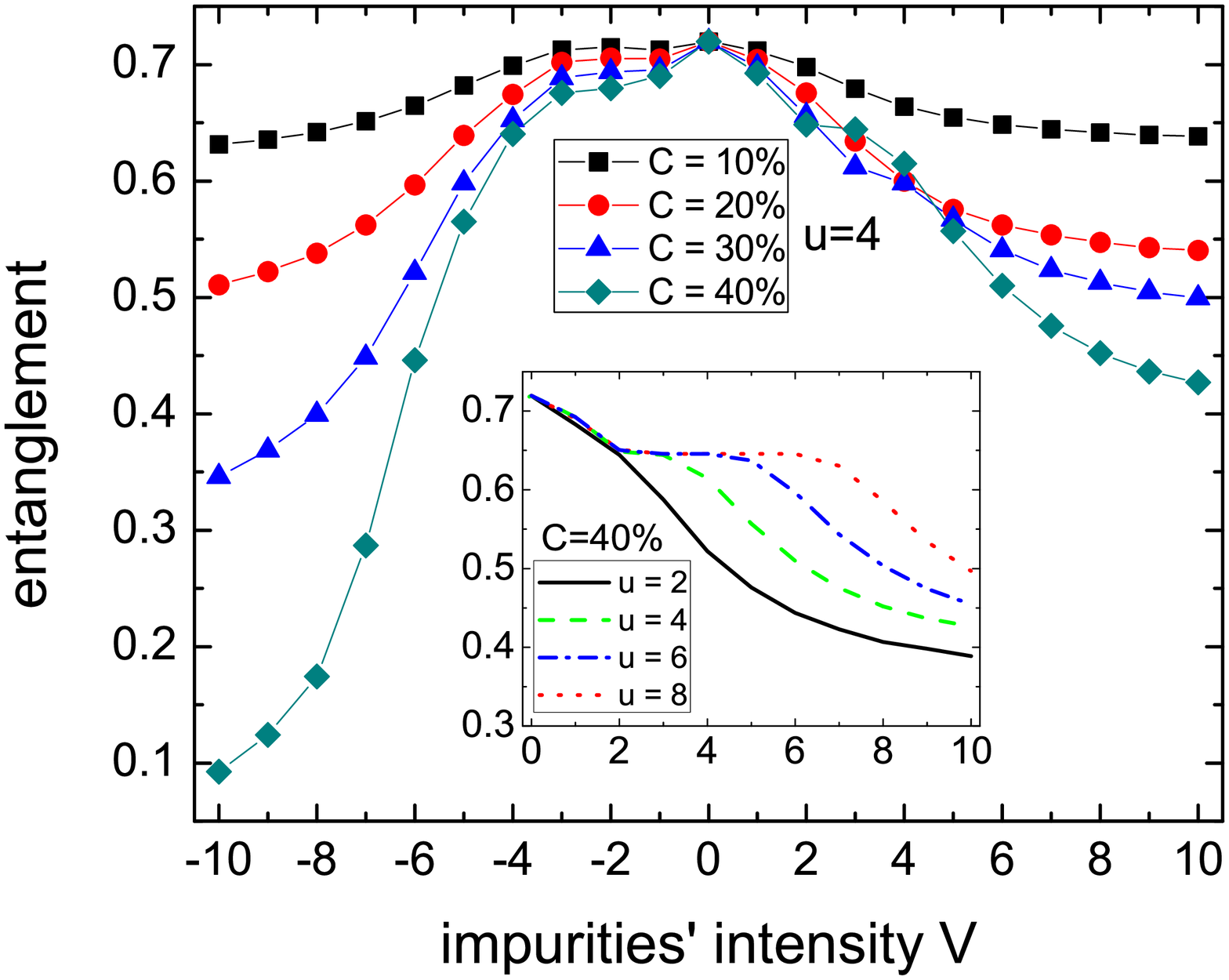}
\includegraphics[width=9cm]{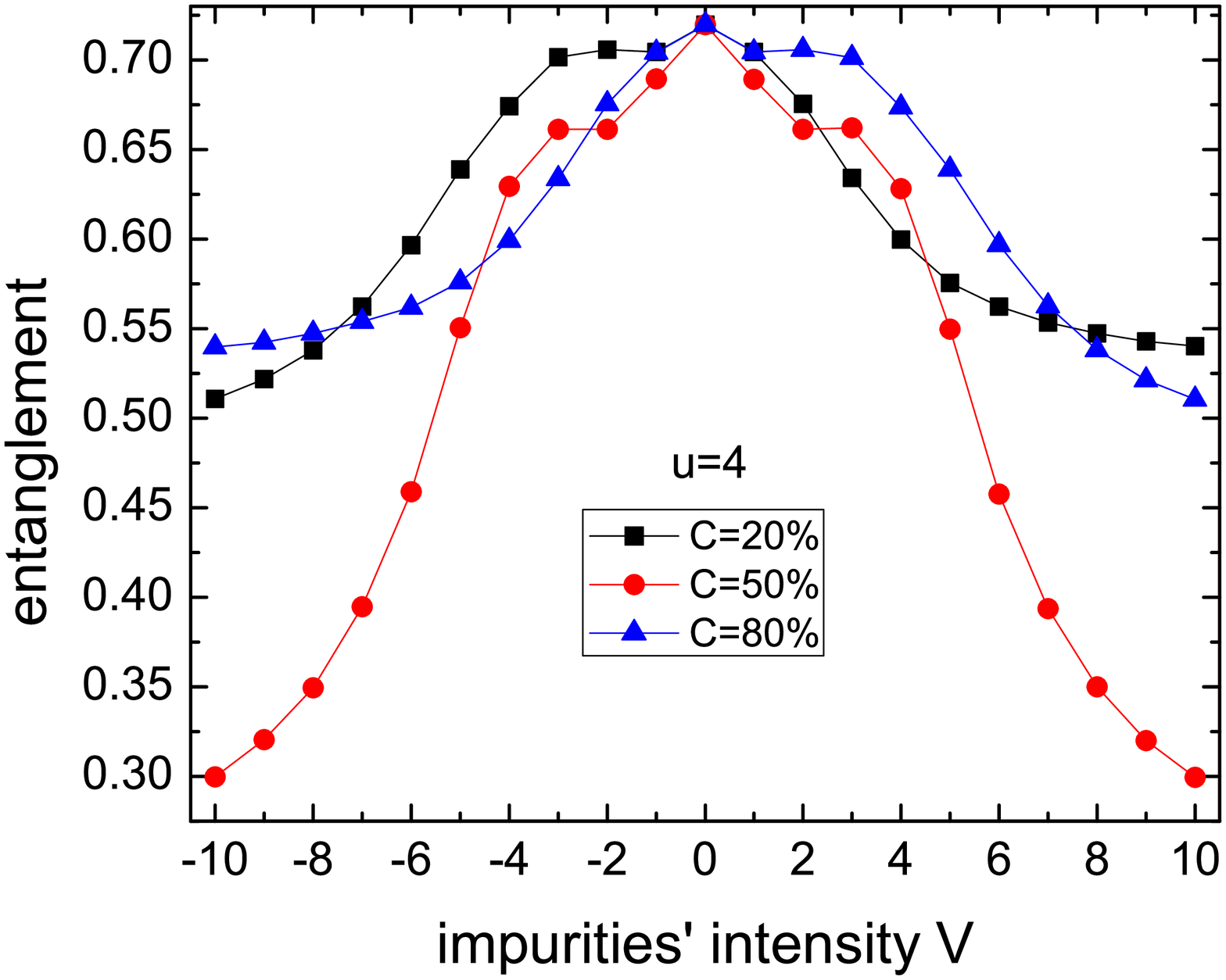}
\caption{Upper panel: average single-site entanglement in disordered chains as a function of the impurities' intensity $V$, for increasing concentrations $C$ of randomly distributed impurities, for 100 sites, 80 particles and $u=4$. $C$ is defined with respect to the total number of sites and each point is an average over 100 samples. {\it Inset}: entanglement as a function of $V$ for $C=40\%$ and different values of $u$. Lower panel: as for the upper panel, but for $C=20\%, 80\%$, and $50\% $, to emphasize the $C\rightarrow 100-C$ symmetry for $V\rightarrow -V$ of the entanglement curves.} \label{disorder}
\end{figure}

We have also checked the performance of our approach for different inhomogeneities and different values of $u$, such as a superlattice defined by a periodic modulation $\Delta V$ in the external potential ($u=3$), and the presence of a {\it single} impurity with intensity $V_{i=51}$ ($u=6$) as shown in Fig. \ref{harm}--b and Fig. \ref{harm}--c, respectively. The agreement is quantitatively similar to that found for the harmonic confinement: for the superlattice the deviations are smaller than $2\%$, while for the impurity system we found deviations of about $2.2\%$ for all impurity intensities (note the scale in the y-axis). 

Our results for three different physical systems confirms that the entanglement can be calculated directly from density measurements by using our density-functional even for inhomogeneous systems. As all the above tests have confirmed the reliability of our functional, we next use our density-functional for investigating disordered systems.

\section{Entanglement in Disordered Systems}

The investigation of any physical property in disordered systems usually requires an average over a large number of samples in order to improve statistics and avoid results peculiar to specific realizations. Thus, for example, for the generation of a single point in a system of 100 sites considering 100 samples, it would be necessary to perform $10^4$ calculations of the desired average single-site property. This represents a serious limiting factor for estimating the entanglement via methods involving the computationally-expensive ground-state energy calculation. In this sense our density-functional prescription represents a great advantage.

Here we consider the average single-site entanglement over 100 realizations of a disordered system defined by a concentration $C$ of randomly distributed impurities with intensity $V$. In Figure \ref{disorder} we present the entanglement obtained with our density functional as a function of $V$. The entanglement is maximum for non disordered chains ($V=0$) and for $V<0$ a {\it plateau} is observed for $C\lesssim 70\%$. This plateau in the entanglement reflects the competition between the repulsive interaction $u$ and the impurities' attractive potential $V$ at the Mott-insulator transition induced by $n_i=1$ at the {\it impurities' sites}. A plateau is also observed in systems with a single impurity (see Fig. \ref{harm}--c), where the percentage drop of entanglement with the impurity strength is however much smaller. Interestingly, for $C\gtrsim30\%$, we also find a plateau for $V>0$ (see Fig..\ref{disorder}, upper panel). The physics behind both plateaus is similar, but for repulsive $V$ and $u$ one needs a higher impurity concentration to reach the Mott-insulator transition, as in this case it is induced at the {\it sites with no impurities}. The reversed role between sites with and without impurities manifests itself in the reflection symmetry with respect to $V=0$ for systems with $C$ and $C'=100-C$, as shown in Fig.\ref{disorder} -- lower panel. The width and height of the plateaus increase towards smaller impurity intensities with increasing (decreasing) impurity concentrations for $V>0$ ($V<0$); the plateaus' width also increases with $u$ as $|V|\lesssim u$, as a result of the competition between the two potentials (see inset of Fig.\ref{disorder}). 
 
In disordered chains, similarly to what has been observed for other inhomogeneous systems, inhomogeneities are in general destructive to the entanglement, and the higher $C$ and $|V|$, the more the entanglement is destroyed. However our calculations show that there are regimes for which the degree of entanglement is {\it almost unaffected}. The entanglement in this region is almost as higher as the homogeneous case, $V=0$. This property could be useful for indicating the regime of parameters for which real-life devices may present a high degree of entanglement despite the presence of impurities. 

\section{Conclusions}

In conclusion we have derived an analytical density functional for the entanglement in the Hubbard model represented by a very simple mathematical expression. We have shown that it reproduces the entanglement {\it quantitatively} in homogeneous and inhomogeneous systems. We tested our functional on very different inhomogeneous systems, i.e. an harmonic confinement, which simulates experiments of trapped atoms in optical lattices, as well as on a superlattice and a single-impurity system. 

We applied our density-functional prescription to the investigation of the entanglement in disordered systems: this analysis would be computationally very expensive -- and as such difficult to achieve -- with other methods. We found that, although in general the disorder destroys the entanglement, there exist a regime of parameters, both for repulsive and attractive impurity potential, for which the degree of entanglement remains almost unaffected. This result is promising as it suggests potential robustness for future device production. 

In addition of allowing simpler numerical calculations and paving the road to analytical analysis of many-body and quantum information properties, our findings are of direct interest for measuring the entanglement from experimental setups. 

\section{Acknowledgements}

We thank E. A. L. Henn for fruitful discussions. VVF acknowledges funding from Brazilian agency CAPES(4101/09-0) and IDA from EPSRC grant EP/F016719/1.


\begin{references}
\bibitem{note_many_body} This is in contrast to the common belief that the strong interaction among particles would destroy local entanglement easily, one of the reasons for which the investigation of single-site, or more generally, local entanglement in many-body systems was somewhat left behind.
\bibitem{foot1}This procedure ensures the continuity of our approximated functional.
\bibitem{foot2} For all calculations open boundary conditions were used and the densities $n_i$ were obtained via DFT with LDA for the exchange-correlation energy.
\end{references}
\end{document}